%% file: main.tex
\documentclass{article} 
\usepackage{style,times}

\input{math_commands.tex}

\usepackage{hyperref}
\usepackage{url}
\usepackage{subfigure}      
\usepackage{graphicx}       
\usepackage{enumitem}       

\title{Constructing and Compressing Global \\ Moment Descriptors from Local Atomic \\ Environments}

\author{Vahe~Gharakhanyan$^{1}$\thanks{Correspondence to \texttt{vg2471@columbia.edu} and \texttt{au2229@columbia.edu}}\ , Max S.~Aalto$^{2,3}$, Aminah~Alsoulah$^{3}$, \textbf{Nongnuch~Artrith}$^{4}$, \textbf{Alexander~Urban}$^{3}$ \\ \\
$^{1}$ Department of Applied Physics and Applied Mathematics, Columbia University, New York, NY, USA \\
$^{2}$ Physics and Informatics Laboratories, NTT Research, Inc., Sunnyvale, CA, USA \\
$^{3}$ Department of Chemical Engineering, Columbia University, New York, NY, USA \\
$^{4}$ Debye Institute for Nanomaterials Science, Utrecht University, Utrecht, Netherlands
}

\finalcopy
\begin{document}

\maketitle

\begin{abstract}
Local atomic environment descriptors (LAEDs) are used in the materials science and chemistry communities, for example, for the development of machine learning interatomic potentials.
Despite the fact that LAEDs have been extensively studied and benchmarked for various applications, global structure descriptors (GSDs), i.e., descriptors for entire molecules or crystal structures, have been mostly developed independently based on other approaches. 
Here, we propose a systematically improvable methodology for constructing a space of representations of GSDs from LAEDs by incorporating statistical information and information about chemical elements.
We apply the method to construct GSDs of varying complexity for lithium thiophosphate structures that are of interest as solid electrolytes and use an information-theoretic approach to obtain an optimally compressed GSD. 
Finally, we report the performance of the compressed GSD for energy prediction tasks. 
\end{abstract}

\section{Introduction}
Local atomic environment descriptors (LAEDs) are widely used in the materials science and chemistry communities, for example, in machine learning interatomic potentials (force fields) and for detecting (dis)similarities between atomic environments \citep{parsaeifard2021assessment, langer2022representations}.
An early example is the atom-centered \emph{symmetry functions} introduced by Behler and Parrinello \citep{behler2007generalized, behler2011atom} to describe the chemical environment of atoms as input to atomic neural networks.
Since then, various other LAED methods have been proposed in the literature and broadly applied to research questions in chemistry and materials science \citep{drautz2019atomic, onat2020sensitivity, musil2021physics, langer2022representations}.
In parallel, global structure descriptors (GSDs) for entire molecules and crystal structures have independently been developed, though, for periodic crystal structures, only a few representations have been proposed \citep{damewood2023representations}. 
So far, less emphasis has been placed on constructing GSDs from LAEDs for learning tasks where the property of interest cannot be intuitively decomposed into atomic contributions, for example, for predicting elastic properties such as the bulk modulus. 
Given that LAEDs have been extensively benchmarked, it would be desirable to leverage this experience for the construction of GSDs as well.

Prior work includes examples of constructing GSDs by evaluating the mean of the LAEDs for a given atomic structure \citep{priedeman2018quantifying, cheng2020mapping}.
Other authors proposed to include the variance in addition to the mean to combine all LAEDs of sites with the same chemical element \citep{guo2022artificial}.
Higher mathematical moments have been used to construct invertible LAEDs with prospective applications for materials discovery through inverse design tasks \citep{uhrin2021through}. 
Here, we build on and extend these ideas to formalize the construction of GSDs by combining chemical information and structural statistics via a moment expansion of the distribution of atomic environment descriptors. 
We then use a recently introduced information-theoretic approach \citep{glielmo2022ranking, darby2022compressing, darby2022tensor} to inspect the relationship of the GSD information content with its complexity and determine the GSDs that offer the optimal compromise between the information content about the atomic system and the descriptor complexity \citep{zeni2021compact, khan2023quantum}. 
Finally, we demonstrate the performance of our proposed descriptors for energy prediction tasks.

\section{Methodology}

We adopt the following notation. The structure data set $\sS$ consists of atomic structures $s$ that in turn contain the sites $\{a_1,  ..., a_{n_{1}}, b_1, ..., b_{n_{2}}, ...\}$ with corresponding element types $\{A, B, ...\}$.
The LAED of site $a$ in structure $s$ is denoted $\sigma_a^{(s)}$ and is a real-valued vector.
Note that the present work is independent of the method that is used to obtain $\sigma_a^{(s)}$.

\subsection{Chemical Element descriptor}

We call \emph{chemical element descriptor} (CED) a representation of all sites of the same chemical element in a given structure. 
A CED can be constructed by combining the information of the LAEDs of all sites of a given element type. 
Simply averaging all LAEDs would potentially result in a significant loss of information.
One way to systematically go beyond the mean LAED is by taking into account the statistics through mathematical moments. 
The first moment of a distribution is the mean, and general moments of order two (variance) and above are given by $\mu_{n}(\sX)= \dfrac{1}{N}\sum\limits_{i = 1}^{N} \left(X_i - \bar{X}\right)^n$.
The mean of the LAEDs thus corresponds to a CED with an inner-moment\footnote{Here, ``inner'' refers to the CED construction and ``outer'' to the GSD construction of section~\ref{sec:GSD}} degree of 1 ($n_{\text{in}}=1$). 
To incorporate higher moments in the CED, we can stack (concatenate) different moment descriptors:
$
    \sigma_{A}^{(s)} = \bigoplus\limits_{i=0}^{n_{\text{in}}}\mu_{i}
    (\{
    \sigma_{a}^{(s)} \mid \forall a \in A,\ A \in s 
    \}),
$
where the $\bigoplus$ operator concatenates its arguments into a single CED vector with dimensions of $d^{(A)} = d^{({a})}\cdot n_{\text{in}}$, where $d^{({a})}$ is the dimension of the LAED.

\subsection{Global structure descriptor}
\label{sec:GSD}

We propose different methods for obtaining a \emph{global structure descriptor} (GSD) from LAEDs and CEDs.
A global structure descriptor can be constructed from:
\begin{itemize}[leftmargin=2em, labelsep=0cm, align=left, noitemsep, topsep=-4pt]
\item[\ \ i:] LAEDs, by finding the moments up to a degree of outer-moment ($n_{\text{out}}$) of the distribution of all (element-weighted) LAEDs, 
\item[\ ii:] CEDs, by stacking CEDs for all chemical elements,
\item[iii:] CEDs, by finding the moments up to a degree of outer-moment of the distribution of (element-weighted) CEDs.
\end{itemize}
%
\begin{equation}
\sigma^{(s)} = \begin{cases}
    \text{\ \ i:}\ \ \ \bigoplus\limits_{i=0}^{n_{\text{out}}}
    \mu_{i}
    (\{
    w_{A}\cdot \sigma_{a}^{(s)} \mid a \in A,\ \forall a \in s\}), & \text{if $n_{\text{out}}\in \{1,2,...\}$ and $n_{\text{in}}=0$}.\\[1.25em]
    \text{ ii:}\ \ \ \bigoplus\limits_{A \in s}
    {
    \sigma_{A}^{(s)}
    }, & \text{if $n_{\text{out}}=0$ and $n_{\text{in}}\in \{1,2,...\}$}.\\[1.25em]
    \text{iii:}\ \ \ \bigoplus\limits_{i=0}^{n_{\text{out}}}
    \mu_{i}
    (\{
    w_{A}\cdot \sigma_{A}^{(s)} \mid \forall A \in s
    \}), & \text{if $n_{\text{out}}$,\ $n_{\text{in}} \in \{1,2,...\}$}.
    \end{cases}
\label{eq:global}
\end{equation}
The element weighting can be turned off by setting $w=1$ for all elements. 
Note that the LAEDs themselves can already incorporate chemical information, for example, through element weightings \citep{artrith2017efficient}.
The dimension of the final global structure descriptor is $d^{(s)} = d^{(a)}\cdot \text{max}(1, n_{\text{out}} )\cdot \text{max}(1, n_{\text{in}})$ for methods i and iii, and, for method ii, is additionally scaled by the number of unique elements in $\sS$.
For multi-element data sets, where not all structures contain all element types, GSD from method ii should be correctly zero-padded to achieve consistent GSD dimensions among all structures.
Besides the three GSD construction methods shown above, we can create new GSDs by simply stacking different GSD constructions discussed (i.e., stacking GSDs from methods iii and i, or descriptors from method i or iii with and without element weightings). 

\subsection{Information content and compression}
\label{rim-srim}
The number of inner and outer moments considered, as well as the number of unique elements in the data set, can increase the GSD dimension. 
We define the complexity of a descriptor as $d^{(s)} / d^{(a)}$.
It is important to quantify the information that each new moment adds about the system and to create methods for determining the best dimension/construction method of the GSD that contains the necessary structural and chemical information.
In other words, we're looking for the optimum combination of inner and outer moments and element weightings that can be obtained by compressing the full GSD with minimal loss of information.

To compress the descriptor, we need to quantify the information content (or loss) between compressed and full descriptors. 
We can choose between information content measures based on distances or ranks, the latter being preferred because it is agnostic to the scaling of the space. 
Here, we use an information imbalance between two descriptors based on ranks (also called \emph{rank information imbalance} - RIM), recently introduced by \citet{glielmo2022ranking}: $\Delta R (\Sigma_{1} \rightarrow \Sigma_{2})= \dfrac{1}{N} \sum\limits_{\mathbf{s_i}, \mathbf{s_j} \in \sS} \left(r_{s_i s_j}^{(\Sigma_2)}\ \mid\ r_{s_i s_j}^{(\Sigma_1)}=1\right)$, where $r_{s_i s_j}^{(\Sigma)}$ is the rank between GSDs of structures $s_i$ and $s_j$ in descriptor space $\Sigma$. 
The rank $r_{s_i s_j}$
is computed by sorting the distances between GSD of $s_i$ and all other structure descriptors from smallest to largest and then finding the corresponding index for $s_j$. 
So, for instance, a rank of 1 means that GSDs of $s_i$ and $s_j$ are nearest neighbors in the given descriptor space. 
To find the rankings, we use Euclidean distance as the distance metric.
So rank information imbalance computes the average rank in descriptor space $\Sigma_2$ for the structures that are nearest neighbors in descriptor space $\Sigma_1$.
If $\Delta R (\Sigma_{1} \rightarrow \Sigma_{2}) \approx 1$, then descriptor $\Sigma_{1}$ is informative of descriptor $\Sigma_{2}$. The higher the imbalance, the less informative $\Sigma_{1}$ becomes for $\Sigma_{2}$.
It is important to note here that the rank imbalance is not symmetric ($\Delta R (\Sigma_{1} \rightarrow \Sigma_{2}) \neq \Delta R (\Sigma_{2} \rightarrow \Sigma_{1})$) because the sets of pairs of nearest neighbors in different descriptor spaces are, almost often, different.
For this reason, similar to \citet{glielmo2022ranking}, we can construct a \emph{symmetric full rank information imbalance} (SRIM) as follows: $\Delta \bar{R} (\Sigma_{1},\Sigma_{2})=
    \dfrac{1}{2} 
    \left[\Delta R (\Sigma_{1} \rightarrow \Sigma_{2}) + \Delta R (\Sigma_{2} \rightarrow \Sigma_{1})\right]$.
Both information imbalance measures discussed have a lower limit of 1, meaning that both spaces are equivalent (all nearest-neighbor pairs in both descriptor spaces are the same), and the upper limit will depend on the number of structures in the data set. 

\section{Results and discussion}

\subsection{Construction of global structure descriptors}

\textbf{Data set:} We used 6055 previously published computationally generated atomic structures of glass-ceramic lithium thiophosphates (LPS) with compositions on or close to the composition line (Li\textsubscript{2}S)\textsubscript{x}(P\textsubscript{2}S\textsubscript{5})\textsubscript{1-x} \citep{guo2022artificial}. 

\textbf{Local atomic environment descriptor:} 
We employed a LAED based on the truncated Chebyshev expansions of the radial and angular distribution functions (RDF and ADF) \citep{artrith2017efficient}.
This LAED method is numerically efficient and has the advantage that its dimension does not increase with the number of chemical elements; it is also invariant with respect to rotations, translations, and atom permutations \citep{artrith2016implementation}. 
For all elements, we used a cutoff of 6.0 Å with expansion order 19 for the RDF and a cutoff of 3.0 Å with expansion order 5 for the ADF. 
The LAED is a stacking of four sets of expansion coefficients: the coefficients of the RDF and ADF with and without element weightings, leading to a total LAED dimension of $2\times{}(19 + 5) = 48$ \citep{guo2022artificial}.
The following element weightings were used: {Li: -1, P: 0, and S: +1}. 

\textbf{Global structure descriptor:} GSDs with all possible combinations of outer and inner moments up to a degree of five, with and without element weightings (using the LAED weightings), were constructed. 
In the following, $\Sigma_{ijk}$ denotes a GSD with outer moments up to a degree of $i$, inner moments up to a degree of $j$, and with ($k=1$) or without ($k=0$) element weightings. 
For example, $\Sigma_{231}$ and $\Sigma_{230}$ represent GSDs with and without element weighting, respectively, that are constructed with outer moments of up to a degree of two (average and variance) and inner moments of up to a degree of three (average, variance, and third moment). 
Note that GSDs of type $\Sigma_{0j1}$ do not exist, according to \eqref{eq:global}. 
Element-weighted and unweighted GSDs with $i\neq{}0$ and identical inner ($i$) and outer ($j$) moments can be combined via concatenation to create new GSDs, $\Sigma_{ij2}$; for example, $\Sigma_{232} = \Sigma_{230} \oplus \Sigma_{231}$.
With such concatenation, we further construct two GSDs that contain the most information about the geometry statistics and chemistry, $\Sigma_{555} = \Sigma_{552} \oplus \Sigma_{502} \oplus \Sigma_{050}$ and $\Sigma_{554} = \Sigma_{552} \oplus \Sigma_{502}$.
$\Sigma_{555}$ and $\Sigma_{554}$ have the dimensions 3600 and 2880, and complexities 75 and 60, respectively. 
Enumerating all distinct combinations of GSDs with moments up to order 5 and their concatenations resulted in a GSD space with 97 distinct representations. 
All descriptors were normalized to remove the scale imbalance from using higher-order moments.

\subsection{Complexity analysis}

By design, most of the constructed GSDs are contained within other GSDs, and most information is contained in $\Sigma_{555}$, followed by $\Sigma_{554}$. 
Here, we aim to identify those lower-dimensional descriptors that exhibit minimal information loss compared to these two references. 
For this purpose, we evaluated the RIMs between all $97\times{}96=9312$ pairs of GSDs. 

\begin{figure}[ht]
\centering
\hfill
\subfigure[]{\includegraphics[width=6.9cm]{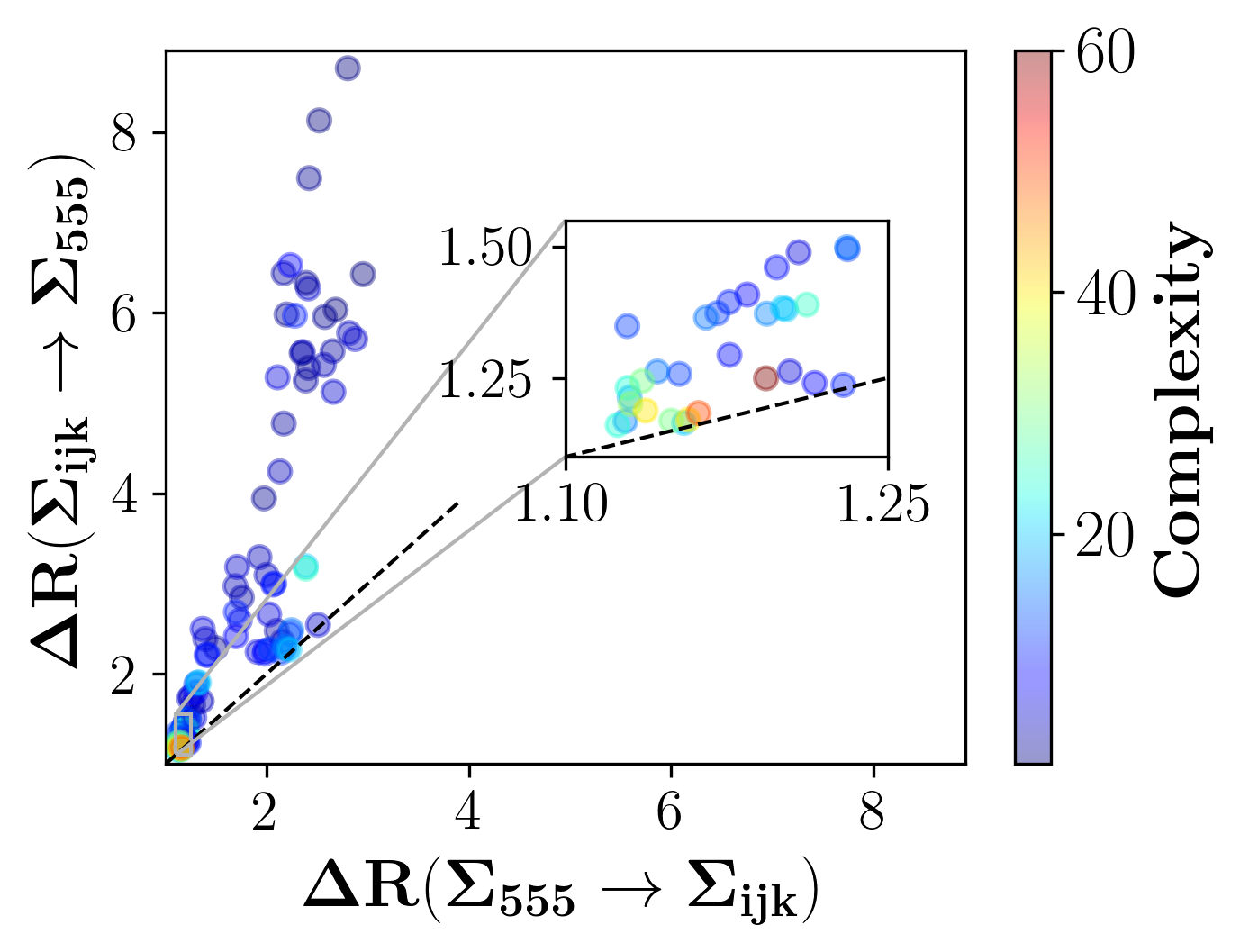}}\label{z}
\hfill
\subfigure[]{\includegraphics[width=6.9cm]{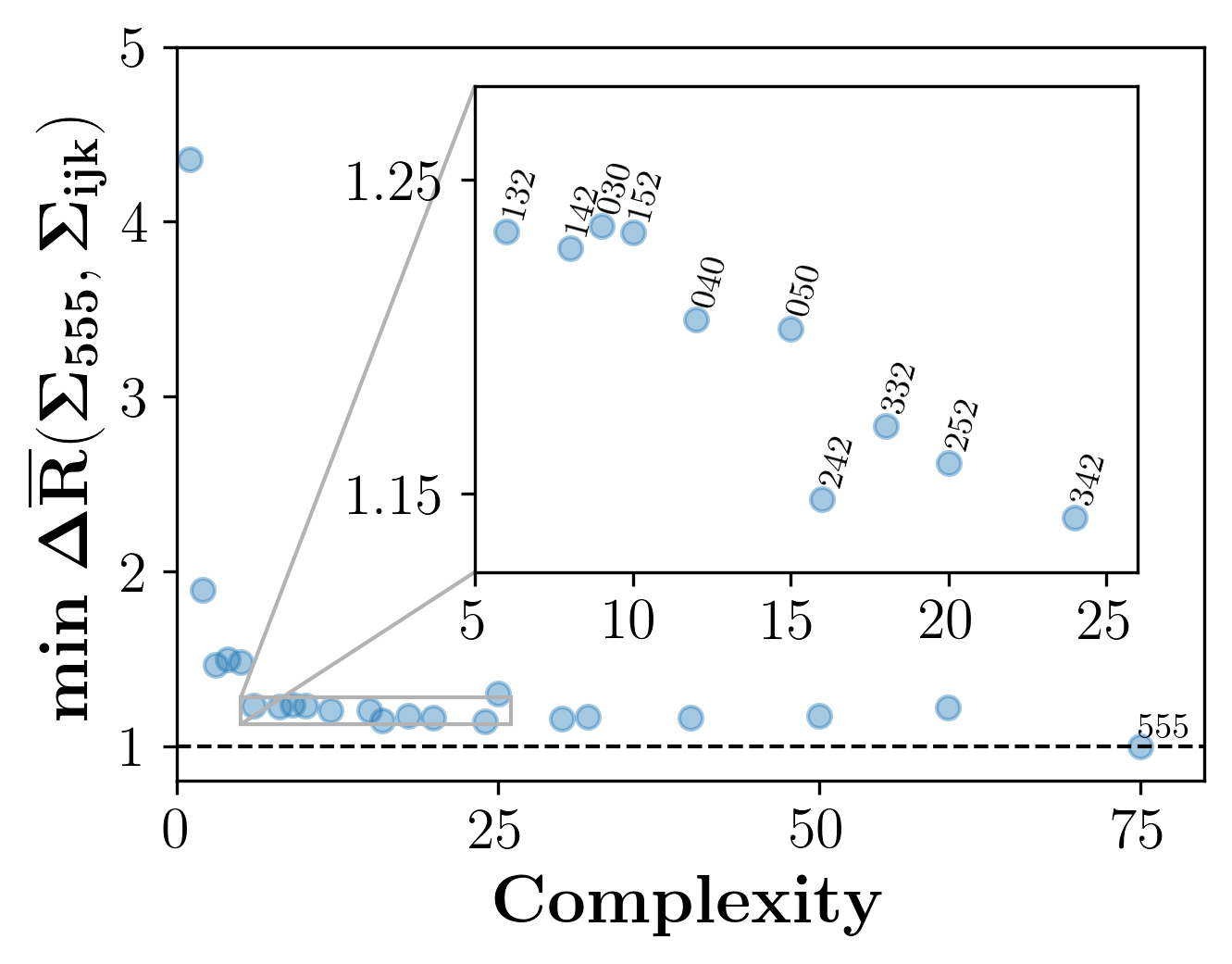}}
\hfill
\caption{
(a) Correlation plot between rank information imbalances of inferring all other descriptors from $\Sigma_{555}$ (on $x$-axis) and inferring $\Sigma_{555}$ from all other descriptors (on $y$-axis). 
(b) Minimum symmetric rank information imbalance between $\Sigma_{555}$ and all other descriptors as a function of the descriptor complexity.
}
\label{fig:all_imbalance}
\end{figure}

Figure~\ref{fig:all_imbalance}a shows a correlation plot between the RIMs for inferring each descriptor space from $\Sigma_{555}$ and inferring $\Sigma_{555}$ from each descriptor space. 
GSDs with high complexity ($>30$) exhibit minor information loss (i.e., retain most information content) when inferring $\Sigma_{555}$ from them or inferring them from $\Sigma_{555}$.
Figure~\ref{fig:all_imbalance}b shows the minimum SRIM between $\Sigma_{555}$ and other descriptors for a given complexity value. 
We observe that increasing the complexity beyond $\Sigma_{242}$ does not affect the SRIM substantially.  
By comparing the results with Figure~\ref{fig:all_no_elems_imbalance}b that shows a similar SRIM analysis for $\Sigma_{554}$, it is apparent that $\Sigma_{i,j,k=2}$ descriptors have the most information content and least information loss for both $\Sigma_{554}$ and $\Sigma_{555}$.
Here we did not mention descriptors $\Sigma_{0,j>2,0}$ because, although their performance towards $\Sigma_{555}$ is good, they provide a worse interpretation for $\Sigma_{554}$, and the reason most likely is that $\Sigma_{554}$ does not contain $\Sigma_{050}$.
The raw data is given in Table~\ref{table-1->} and exhibits the same trend: $\Sigma_{i,j,k=2}$ descriptors are most informative about the other GSDs (see also Table~\ref{table-1->2}) and can be inferred best from other descriptors (Table~\ref{table-->2}). 
This means incorporating only structural/geometric information (with $\Sigma_{ij0}$-type descriptors) is insufficient, and the addition of chemical information to the descriptor through element weightings adds the missing information content.
Table~\ref{table-1->2} also shows that simple averaging of all LAEDs with or without element weighting ($\Sigma_{101}$ and $\Sigma_{100}$, respectively) has the worst overall performance. 
Such descriptors might be useful if all atomic environments are similar in composition, but if the environments vary drastically (Figure~\ref{fig:mean_values}), simple averaging and disregarding chemical information leads to information loss.
While $\Sigma_{101}$ and $\Sigma_{100}$ are the worst for inferring other GSDs, the data in Table~\ref{table-->2} shows that it is the hardest to infer $\Sigma_{050}$ from other GSDs. 
This is intuitive because $\Sigma_{i>0,j,k\in\{0,1\}}$ GSDs lose either the distinct elemental information by including only element-weight agnostic moments or lose the distinct geometry information by including element-weighted moments.
From Table~\ref{table-most_similar_pairs}, we see that GSDs that are most similar to each other contain higher moments and the same method for the descriptor construction.
This exemplifies that adding the next higher moment to the descriptor results in a decreasing information gain, i.e., the GSDs converge with the order of the moment expansion.
From Table~\ref{table-most_dissimilar_pairs}, the most dissimilar descriptors are those constructed with different methods, meaning that different construction methods will result in including different pieces of information.

\subsection{Energy fittings}

In the previous section, we assessed the information content in different GSDs.
Here, we investigate whether high information content is indeed beneficial for learning tasks.
We trained on a subset of formation energies of those structures in the LPS dataset that are exactly on the (Li\textsubscript{2}S)\textsubscript{x}(P\textsubscript{2}S\textsubscript{5})\textsubscript{1-x} composition line \citep{guo2022artificial}. 
We used a gradient-boosted tree model with four different numbers of estimators to accommodate the varying complexity of GSDs. 
Training and test set splits of 9 to 1 over five random seed numbers were used (Figure~\ref{fig:energies}). 
From the mean absolute error (MAE) and root mean squared error (RMSE) plots, it can be seen that indeed $\Sigma_{100}$ and $\Sigma_{101}$ are performing the worst towards the energy training task. 
$\Sigma_{i,j,k=2}$ descriptors perform better, although they do not reach the accuracy of the full $\Sigma_{555}$ descriptor. 
Interestingly, the descriptor that, on average, contained most information about other descriptors (Table~\ref{table-1->}), $\Sigma_{332}$, is performing second best after the full descriptor and performs equivalently when considering the MAE only. 
Overall, the scale of energy errors is on the same order of magnitude as the state-of-the-art neural network interatomic potentials, even though the tree models and LAED parameters were not optimized.
Important to observe that the relative rankings of optimal GSDs from the information-theoretic approach and energy models are slightly different. 
One reason is that the models were trained on a subset of LPS structures (about 2/3 of the original data set) and evaluated on a test set that contains only 7\% of all structures. 
To conclude, the results of the information-theoretic approach and energy models are sensitive to the data set used. 
A similar analysis should first be performed for a new data set before choosing an optimally compressed descriptor. 
See Appendix~\ref{appendix:d} and~\ref{appendix:e} for further discussion.
This analysis leads to two conclusions: (i)~it is apparent that the complexity of the full descriptor is not needed, and models with similar or equivalent performance can be obtained with less complex GSDs,
(ii)~the systematic convergence of the GSDs seen in the information imbalance is also reflected by the performance of the energy models. 

\section{Conclusion}

We introduced a systematic framework for constructing computationally efficient and property-independent global structure descriptors from local atomic environment descriptors by incorporating both geometry statistics through mathematical moments and chemistry through element weightings. 
We demonstrated for a set of glassy/amorphous lithium thiophosphate structures how global descriptors with an optimal balance of information content and complexity can be identified, and we confirmed the descriptor performance with an energy prediction task.
In future work, we plan to investigate how the hyperparameters of the local atomic environment descriptor affect the performance of the global descriptor, how the descriptors perform for different moment expansions (through standardized moments or cumulants) and on much larger materials data sets with diverse chemical elements.

\section*{Code availability}
The code to construct global structure descriptors from local environment descriptors is implemented in \href{http://ann.atomistic.net/}{\color{blue}ænet} software package and can be found in \href{https://github.com/atomisticnet/aenet-python}{\color{blue}https://github.com/atomisticnet/aenet-python} open source repository.

\section*{Acknowledgements}
This work was supported by the National Science Foundation under Grant No. DMR-1940290 (Harnessing the Data Revolution, HDR).
The authors thank Dallas R. Trinkle and Snigdhansu Chatterjee for helpful discussions.

\bibliography{bibliography}
\bibliographystyle{bibliography_style}

\newpage
\appendix
\counterwithin{figure}{section}
\counterwithin{table}{section}

\section{Appendix: Statistics of individual components of the full global structure descriptor}
\label{appendix:a}

\begin{figure}[ht]
\centering
\subfigure[]{\includegraphics[width=13.8cm]{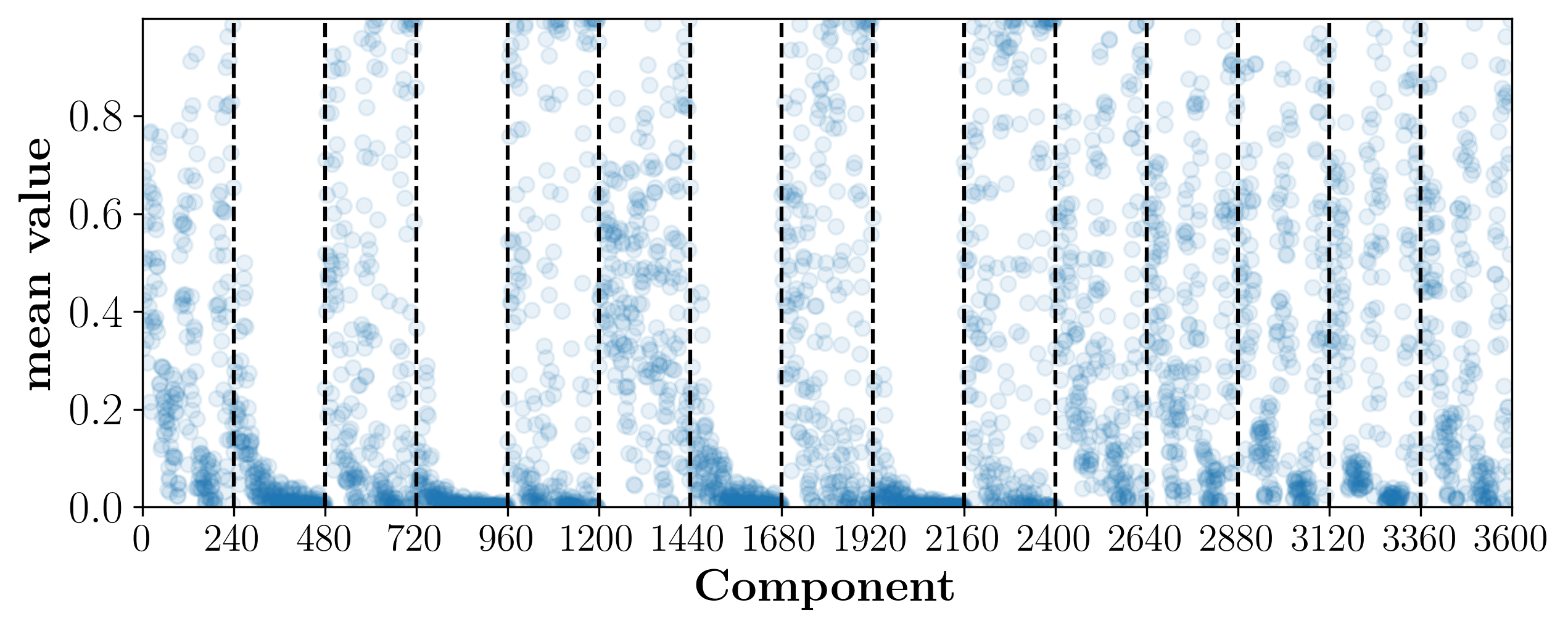}}
\subfigure[]{\includegraphics[width=13.8cm]{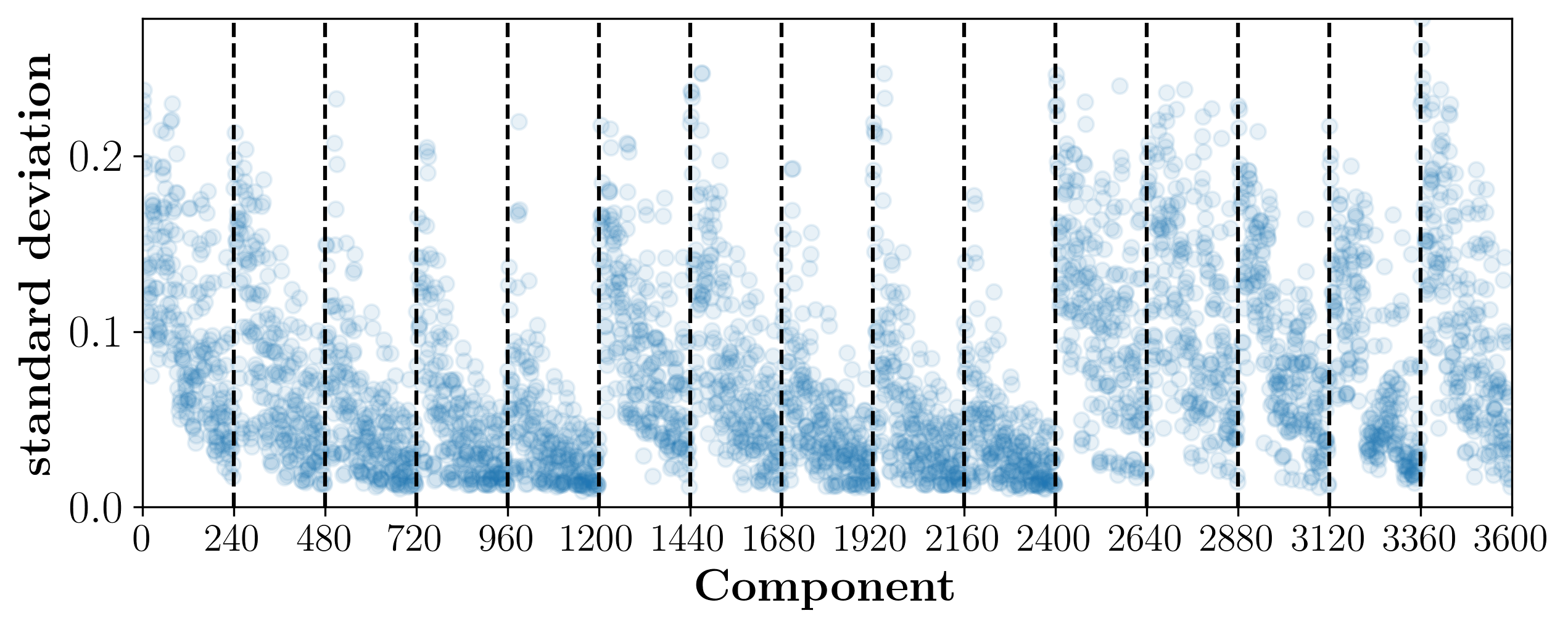}}
\caption{
(a) Mean and
(b) standard deviation values of each component in $\Sigma_{555}$ global structure descriptor. 
%
}

\label{fig:mean_values}
\end{figure}

\newpage
\section{Appendix: Additional information imbalance analysis}
\label{appendix:b}

\begin{figure}[ht]
\centering
\hfill
\subfigure[]{\includegraphics[width=6.9cm]{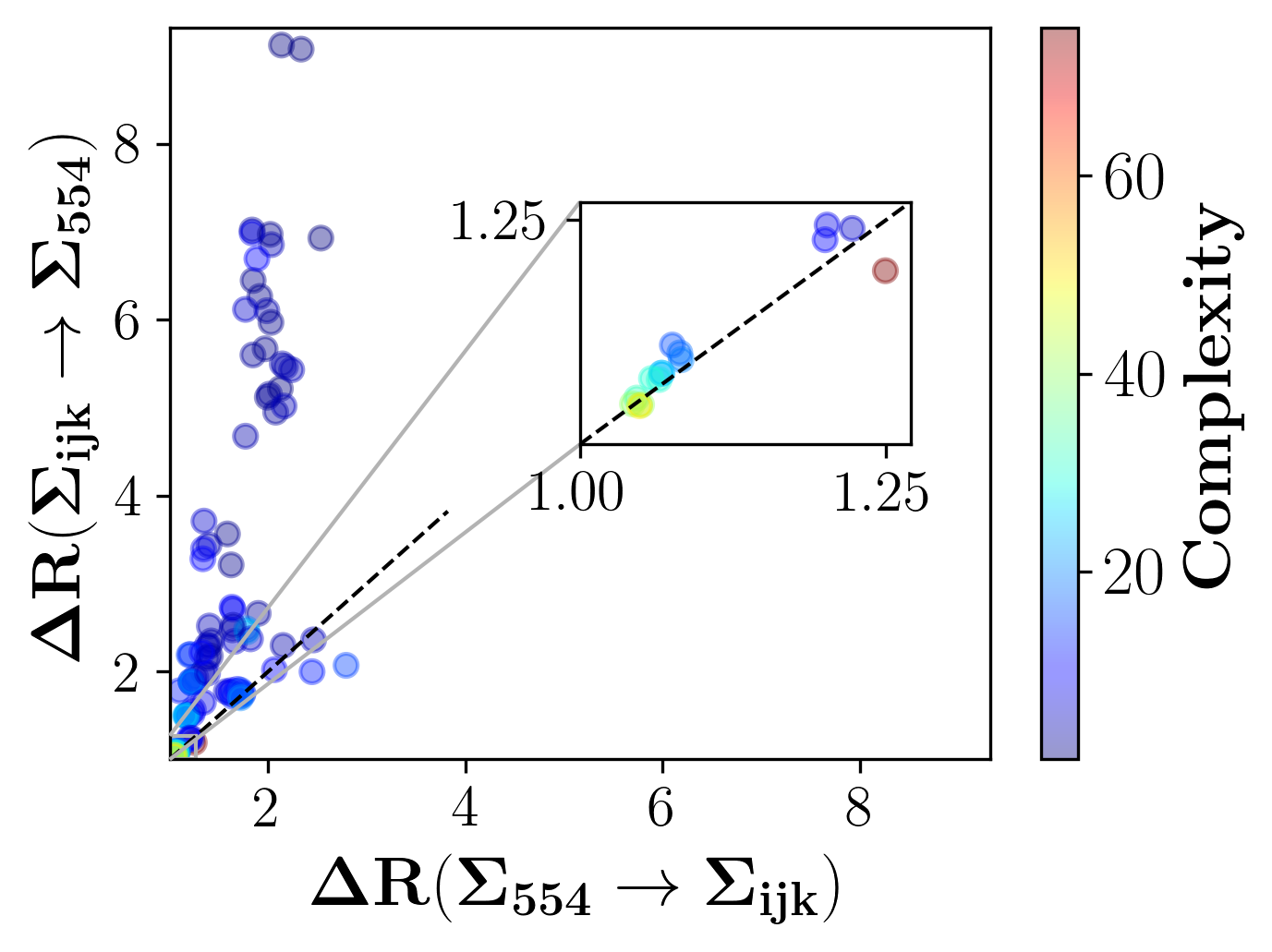}}
\hfill
\subfigure[]{\includegraphics[width=6.9cm]{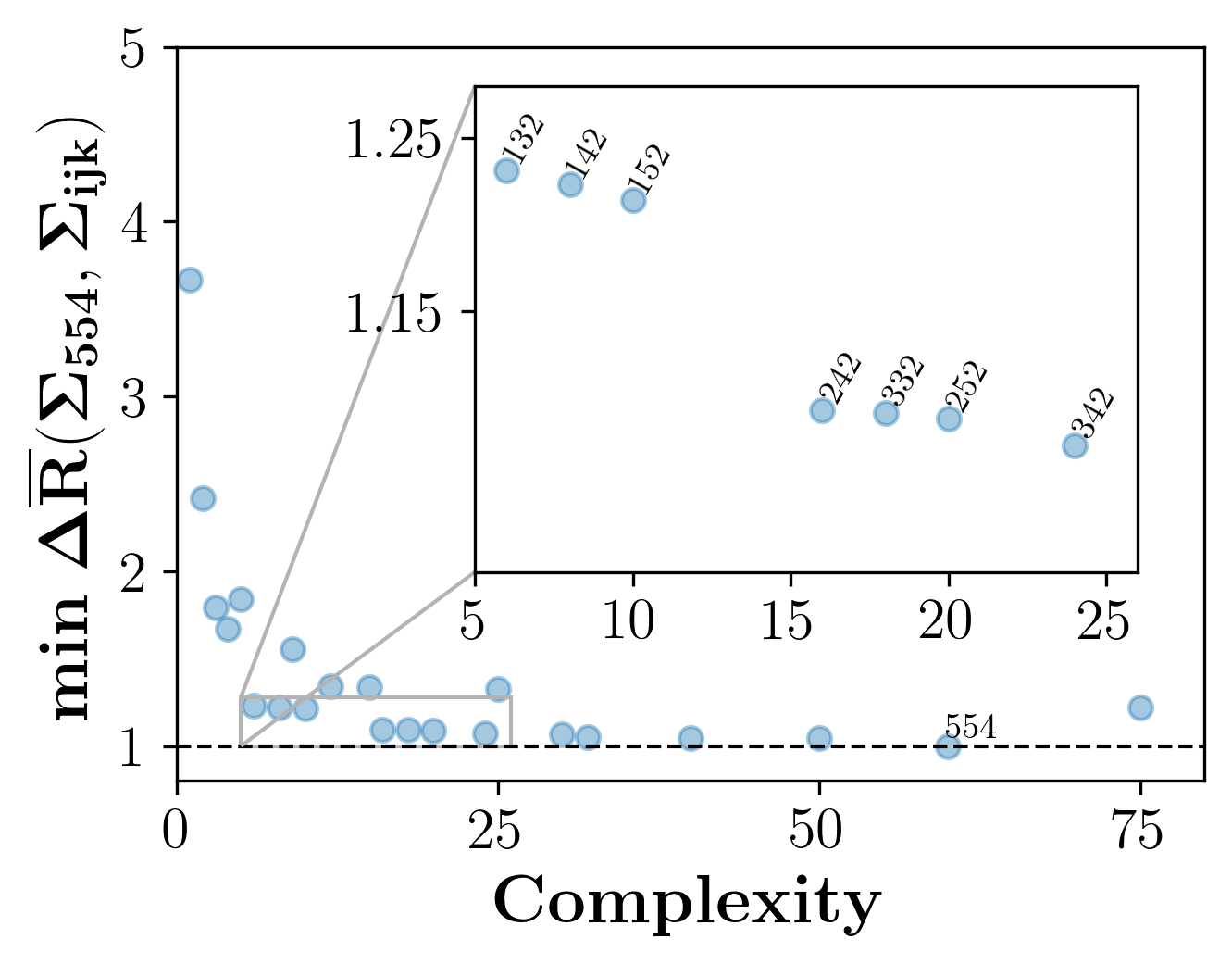}}
\hfill
\caption{
(a) Correlation plot between rank information imbalances of inferring all other descriptors from $\Sigma_{554}$ (on $x$-axis) and inferring $\Sigma_{554}$ from all other descriptors (on $y$-axis). 
(b) Minimum symmetric rank information imbalance between $\Sigma_{554}$ and all other descriptors as a function of the descriptor complexity.
}
\label{fig:all_no_elems_imbalance}
\end{figure}

\begin{figure}[ht]
\centering
\hfill
\subfigure[]{\includegraphics[width=6.72cm]{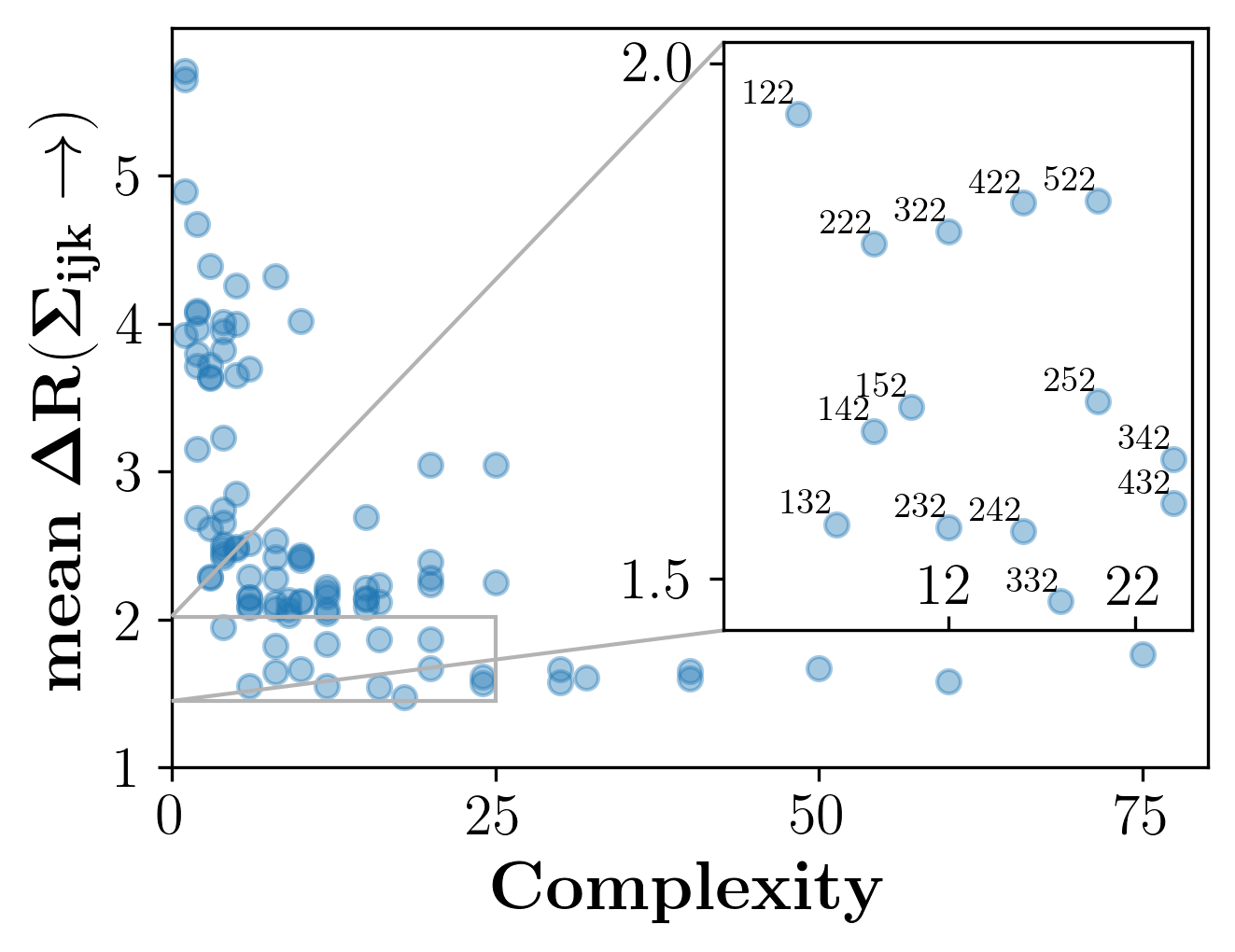}}
\hfill
\subfigure[]{\includegraphics[width=7.08cm]{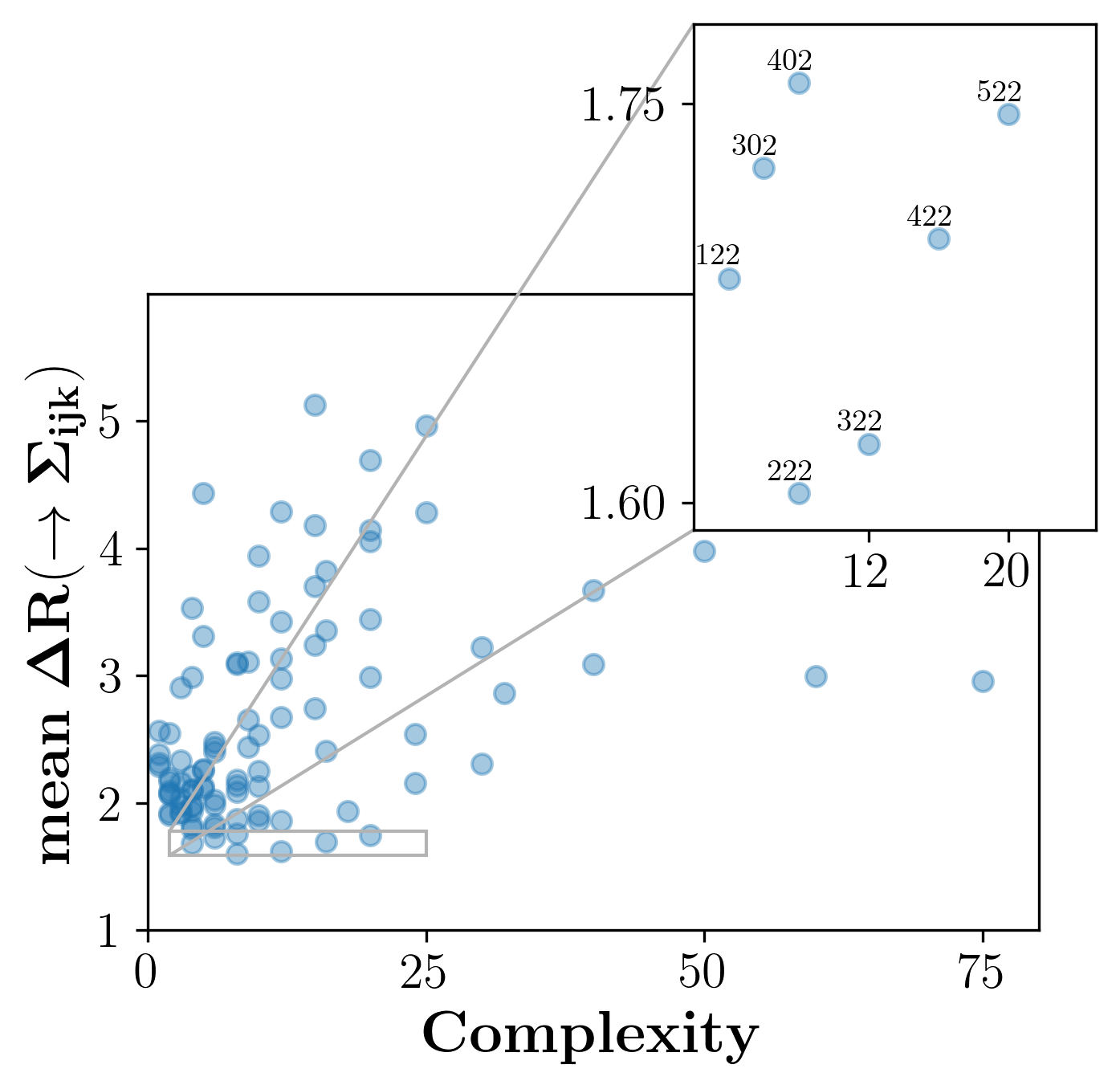}}
\hfill
\caption{
(a) Mean rank information imbalance of inferring all other descriptors from the given descriptor with the given complexity.
(b) Mean information imbalance of inferring the given descriptor with the given complexity from all other descriptors.
}
\label{fig:mean_imbalances}
\end{figure}

\begin{table}[ht]
\label{table:best_ones}
\caption{Global descriptors sorted by the average of symmetric rank information imbalance $\left(\left\langle\Delta \bar{R}\right\rangle\right)$.}
\label{table-1->2}
\begin{center}
    \begin{tabular}{cccc}
        \multicolumn{1}{c}{\bf $\Sigma_1$} & 
        \multicolumn{1}{c}{$\left\langle\Delta R_{1\rightarrow2}\right\rangle$} & 
        \multicolumn{1}{c}{$\left\langle\Delta R_{2\rightarrow1}\right\rangle$} & \multicolumn{1}{c}{$\left\langle\Delta \bar{R}\right\rangle$} 
        \\ \hline
        132 & 1.553 & 1.807 & 1.680 \\ 
        232 & 1.550 & 1.862 & 1.706 \\ 
        332 & 1.478 & 1.938 & 1.708 \\ 
        222 & 1.825 & 1.604 & 1.714 \\ 
        322 & 1.837 & 1.622 & 1.730 \\ 
        422 & 1.864 & 1.699 & 1.782 \\ 
        522 & 1.867 & 1.746 & 1.806 \\ 
        122 & 1.950 & 1.684 & 1.817 \\ 
        432 & 1.573 & 2.158 & 1.865 \\ 
        302 & 2.083 & 1.726 & 1.904 \\ 
        142 & 1.643 & 2.182 & 1.913 \\ 
        402 & 2.077 & 1.758 & 1.918 \\ 
        532 & 1.577 & 2.310 & 1.944 \\ 
        242 & 1.546 & 2.411 & 1.978 \\ 
        502 & 2.126 & 1.858 & 1.992 \\ 
        ... & ... & ... & ... \\ 
        010 & 4.389 & 1.928 & 3.159 \\ 
        040 & 2.045 & 4.291 & 3.168 \\ 
        251 & 2.402 & 3.940 & 3.171 \\ 
        510 & 4.255 & 2.109 & 3.182 \\ 
        541 & 2.282 & 4.147 & 3.214 \\ 
        550 & 2.249 & 4.283 & 3.266 \\ 
        102 & 4.671 & 2.172 & 3.421 \\ 
        351 & 2.693 & 4.183 & 3.438 \\ 
        050 & 2.136 & 5.127 & 3.631 \\ 
        151 & 2.852 & 4.433 & 3.643 \\ 
        110 & 4.891 & 2.564 & 3.728 \\ 
        451 & 3.049 & 4.691 & 3.870 \\ 
        551 & 3.043 & 4.961 & 4.002 \\ 
        100 & 5.710 & 2.314 & 4.012 \\ 
        101 & 5.651 & 2.384 & 4.017 \\ 
\end{tabular}
\end{center}
\end{table}

\begin{table}[ht]
\label{table:best_1}
\caption{Global descriptors sorted by the average of information they contain about all other descriptors $\left(\left\langle\Delta R_{1\rightarrow2}\right\rangle\right)$.}
\label{table-1->}
\begin{center}
    \begin{tabular}{cccc}
        \multicolumn{1}{c}{\bf $\Sigma_1$} & 
        \multicolumn{1}{c}{$\left\langle\Delta R_{1\rightarrow2}\right\rangle$} & 
        \multicolumn{1}{c}{$\left\langle\Delta R_{2\rightarrow1}\right\rangle$} & \multicolumn{1}{c}{$\left\langle\Delta \bar{R}\right\rangle$} 
        \\ \hline
        332 & 1.478 & 1.938 & 1.708 \\ 
        242 & 1.546 & 2.411 & 1.978 \\ 
        232 & 1.550 & 1.862 & 1.706 \\ 
        132 & 1.553 & 1.807 & 1.680 \\ 
        432 & 1.573 & 2.158 & 1.865 \\ 
        532 & 1.577 & 2.310 & 1.944 \\ 
        554 & 1.586 & 2.998 & 2.292 \\ 
        542 & 1.604 & 3.088 & 2.346 \\ 
        442 & 1.605 & 2.861 & 2.233 \\ 
        342 & 1.616 & 2.540 & 2.078 \\ 
        142 & 1.643 & 2.182 & 1.913 \\ 
        452 & 1.651 & 3.670 & 2.660 \\ 
        352 & 1.662 & 3.226 & 2.444 \\ 
        152 & 1.666 & 2.538 & 2.102 \\ 
        252 & 1.672 & 2.986 & 2.329 \\ 
        ... & ... & ... & ... \\             
        111 & 3.923 & 2.291 & 3.107 \\ 
        410 & 3.949 & 2.097 & 3.023 \\ 
        210 & 3.965 & 2.070 & 3.018 \\ 
        511 & 3.998 & 2.258 & 3.128 \\ 
        411 & 4.011 & 2.212 & 3.112 \\ 
        512 & 4.015 & 1.906 & 2.961 \\ 
        112 & 4.075 & 1.924 & 2.999 \\ 
        201 & 4.089 & 2.073 & 3.081 \\ 
        510 & 4.255 & 2.109 & 3.182 \\ 
        412 & 4.317 & 1.875 & 3.096 \\ 
        010 & 4.389 & 1.928 & 3.159 \\ 
        102 & 4.671 & 2.172 & 3.421 \\ 
        110 & 4.891 & 2.564 & 3.728 \\ 
        101 & 5.651 & 2.384 & 4.017 \\ 
        100 & 5.710 & 2.314 & 4.012 \\ 
\end{tabular}
\end{center}
\end{table}

\begin{table}[ht]
\label{table:best_2}
\caption{Global descriptors sorted by the average of information all other descriptors contain about them  $\left(\left\langle\Delta R_{1\rightarrow2}\right\rangle\right)$.}
\label{table-->2}
\begin{center}
    \begin{tabular}{cccc}
        \multicolumn{1}{c}{\bf $\Sigma_2$} & 
        \multicolumn{1}{c}{$\left\langle\Delta R_{1\rightarrow2}\right\rangle$} & 
        \multicolumn{1}{c}{$\left\langle\Delta R_{2\rightarrow1}\right\rangle$} & \multicolumn{1}{c}{$\left\langle\Delta \bar{R}\right\rangle$} 
        \\ \hline
        222 & 1.604 & 1.825 & 1.714 \\ 
        322 & 1.622 & 1.837 & 1.730 \\ 
        122 & 1.684 & 1.950 & 1.817 \\ 
        422 & 1.699 & 1.864 & 1.782 \\ 
        302 & 1.726 & 2.083 & 1.904 \\ 
        522 & 1.746 & 1.867 & 1.806 \\ 
        402 & 1.758 & 2.077 & 1.918 \\ 
        202 & 1.800 & 2.742 & 2.271 \\ 
        132 & 1.807 & 1.553 & 1.680 \\ 
        212 & 1.824 & 3.227 & 2.526 \\ 
        312 & 1.833 & 3.694 & 2.764 \\ 
        502 & 1.858 & 2.126 & 1.992 \\ 
        232 & 1.862 & 1.550 & 1.706 \\ 
        412 & 1.875 & 4.317 & 3.096 \\ 
        200 & 1.905 & 3.155 & 2.530 \\
        ... & ... & ... & ... \\  
        250 & 3.583 & 2.120 & 2.851 \\ 
        452 & 3.670 & 1.651 & 2.660 \\ 
        350 & 3.703 & 2.085 & 2.894 \\ 
        441 & 3.821 & 2.231 & 3.026 \\ 
        251 & 3.940 & 2.402 & 3.171 \\ 
        552 & 3.979 & 1.674 & 2.827 \\ 
        450 & 4.057 & 2.239 & 3.148 \\ 
        541 & 4.147 & 2.282 & 3.214 \\ 
        351 & 4.183 & 2.693 & 3.438 \\ 
        550 & 4.283 & 2.249 & 3.266 \\ 
        040 & 4.291 & 2.045 & 3.168 \\ 
        151 & 4.433 & 2.852 & 3.643 \\ 
        451 & 4.691 & 3.049 & 3.870 \\ 
        551 & 4.961 & 3.043 & 4.002 \\ 
        050 & 5.127 & 2.136 & 3.631 \\ 
\end{tabular}
\end{center}
\end{table}

\begin{table}[ht]
\label{table:most_similar}
\caption{Most similar pairs of descriptors according to the symmetric rank information imbalance $\left(\Delta \bar{R}\right)$. Note that not all decimal points are shown here.}
\label{table-most_similar_pairs}
\begin{center}
    \begin{tabular}{ccccc}
        \multicolumn{1}{c}{\bf $\Sigma_1$} & 
        \multicolumn{1}{c}{\bf $\Sigma_2$} & 
        \multicolumn{1}{c}{$\Delta R_{1\rightarrow2}$} & \multicolumn{1}{c}{$\Delta R_{2\rightarrow1}$} & \multicolumn{1}{c}{$\Delta \bar{R}$} 
        \\ \hline
        450 & 550 & 1.010 & 1.009 & 1.010 \\ 
        420 & 520 & 1.011 & 1.011 & 1.011 \\ 
        421 & 521 & 1.011 & 1.011 & 1.011 \\ 
        440 & 540 & 1.012 & 1.011 & 1.011 \\ 
        442 & 542 & 1.012 & 1.011 & 1.011 \\ 
        411 & 511 & 1.012 & 1.012 & 1.012 \\ 
        432 & 532 & 1.012 & 1.011 & 1.012 \\ 
        451 & 551 & 1.013 & 1.012 & 1.012 \\ 
        422 & 522 & 1.012 & 1.012 & 1.012 \\ 
        431 & 531 & 1.012 & 1.013 & 1.012 \\ 
        452 & 552 & 1.013 & 1.012 & 1.012 \\ 
        430 & 530 & 1.014 & 1.012 & 1.013 \\ 
        412 & 512 & 1.014 & 1.013 & 1.013 \\ 
        441 & 541 & 1.015 & 1.013 & 1.014 \\ 
        401 & 501 & 1.017 & 1.017 & 1.017 \\ 
        340 & 440 & 1.018 & 1.018 & 1.018 \\ 
        311 & 411 & 1.020 & 1.019 & 1.019 \\ 
        410 & 510 & 1.022 & 1.019 & 1.020 \\ 
        352 & 452 & 1.021 & 1.020 & 1.020 \\ 
        330 & 430 & 1.022 & 1.020 & 1.021 \\ 
        252 & 352 & 1.021 & 1.021 & 1.021 \\ 
        341 & 441 & 1.025 & 1.019 & 1.022 \\ 
        320 & 420 & 1.023 & 1.021 & 1.022 \\ 
        340 & 350 & 1.023 & 1.021 & 1.022 \\ 
        342 & 442 & 1.020 & 1.025 & 1.022 \\ 
        232 & 332 & 1.024 & 1.021 & 1.023 \\ 
        351 & 451 & 1.023 & 1.022 & 1.023 \\ 
        140 & 150 & 1.024 & 1.021 & 1.023 \\ 
        251 & 351 & 1.023 & 1.022 & 1.023 \\ 
        322 & 422 & 1.023 & 1.023 & 1.023 \\
        ... & ... & ... & ... & ... \\        
\end{tabular}
\end{center}
\end{table}

\begin{table}[ht]
\label{table:most_dissimilar}
\caption{Most dissimilar pairs of descriptors according to the symmetric rank information imbalance $\left(\Delta \bar{R}\right)$.}
\label{table-most_dissimilar_pairs}
\begin{center}
    \begin{tabular}{ccccc}
        \multicolumn{1}{c}{\bf $\Sigma_1$} & 
        \multicolumn{1}{c}{\bf $\Sigma_2$} & 
        \multicolumn{1}{c}{$\Delta R_{1\rightarrow2}$} & \multicolumn{1}{c}{$\Delta R_{2\rightarrow1}$} & \multicolumn{1}{c}{$\Delta \bar{R}$} 
        \\ \hline
        ... & ... & ... & ... & ... \\        
        452 & 101 & 2.467 & 11.708 & 7.088 \\ 
        441 & 100 & 2.310 & 11.897 & 7.104 \\ 
        452 & 100 & 2.590 & 11.701 & 7.145 \\ 
        112 & 551 & 11.392 & 2.961 & 7.176 \\ 
        251 & 101 & 2.562 & 11.889 & 7.225 \\ 
        450 & 101 & 3.987 & 10.571 & 7.279 \\ 
        102 & 451 & 11.475 & 3.088 & 7.282 \\ 
        550 & 100 & 3.669 & 10.898 & 7.283 \\ 
        102 & 050 & 11.772 & 2.956 & 7.364 \\ 
        151 & 101 & 3.195 & 11.574 & 7.384 \\ 
        552 & 100 & 2.581 & 12.383 & 7.482 \\ 
        541 & 101 & 2.237 & 12.750 & 7.493 \\ 
        541 & 100 & 2.305 & 12.729 & 7.517 \\ 
        102 & 551 & 12.050 & 3.044 & 7.547 \\ 
        550 & 101 & 4.002 & 11.094 & 7.548 \\ 
        552 & 101 & 2.445 & 12.721 & 7.583 \\ 
        110 & 351 & 10.666 & 4.502 & 7.584 \\ 
        111 & 050 & 11.867 & 3.381 & 7.624 \\ 
        100 & 050 & 12.164 & 3.132 & 7.648 \\ 
        251 & 100 & 2.720 & 12.647 & 7.683 \\ 
        151 & 100 & 3.413 & 12.272 & 7.843 \\ 
        101 & 050 & 12.383 & 3.340 & 7.862 \\ 
        351 & 101 & 2.902 & 13.145 & 8.023 \\ 
        110 & 451 & 12.108 & 4.468 & 8.288 \\ 
        351 & 100 & 3.442 & 13.150 & 8.296 \\ 
        110 & 551 & 12.544 & 4.392 & 8.468 \\ 
        451 & 101 & 2.978 & 14.054 & 8.516 \\ 
        551 & 101 & 2.925 & 15.045 & 8.985 \\ 
        451 & 100 & 3.518 & 14.626 & 9.072 \\ 
        551 & 100 & 3.469 & 15.415 & 9.442 \\ 
\end{tabular}
\end{center}
\end{table}

\clearpage
\section{Appendix: A gradient-boosted tree model for formation energy predictions}
\label{appendix:c}

We used a gradient-boosted tree model to determine if our results can compare to those from the state-of-the-art neural network interatomic potential models.

\begin{figure}[ht]
\centering
\hfill
\subfigure[]{\includegraphics[width=6.30cm]{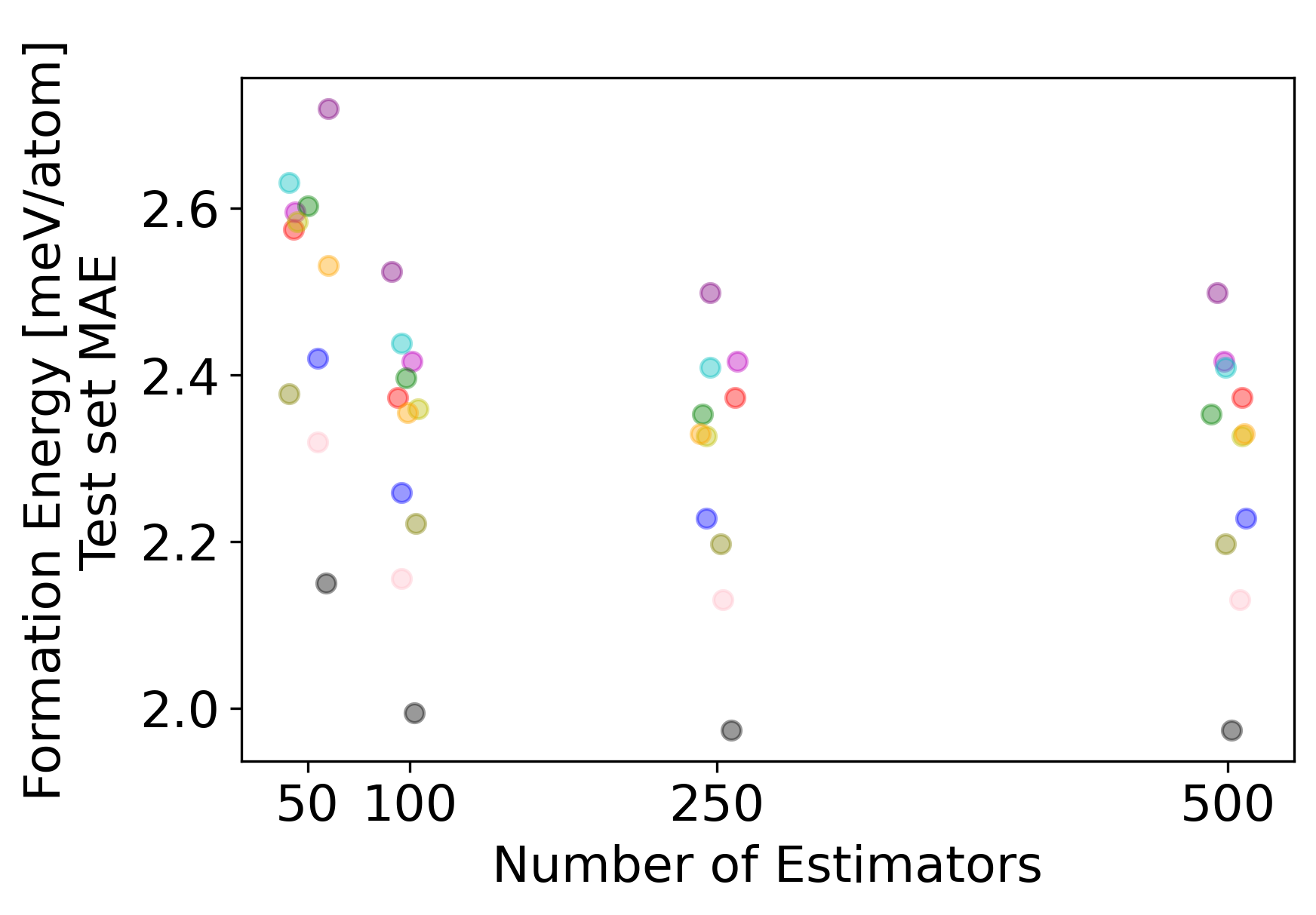}}
\hfill
\subfigure[]{\includegraphics[width=7.54cm]{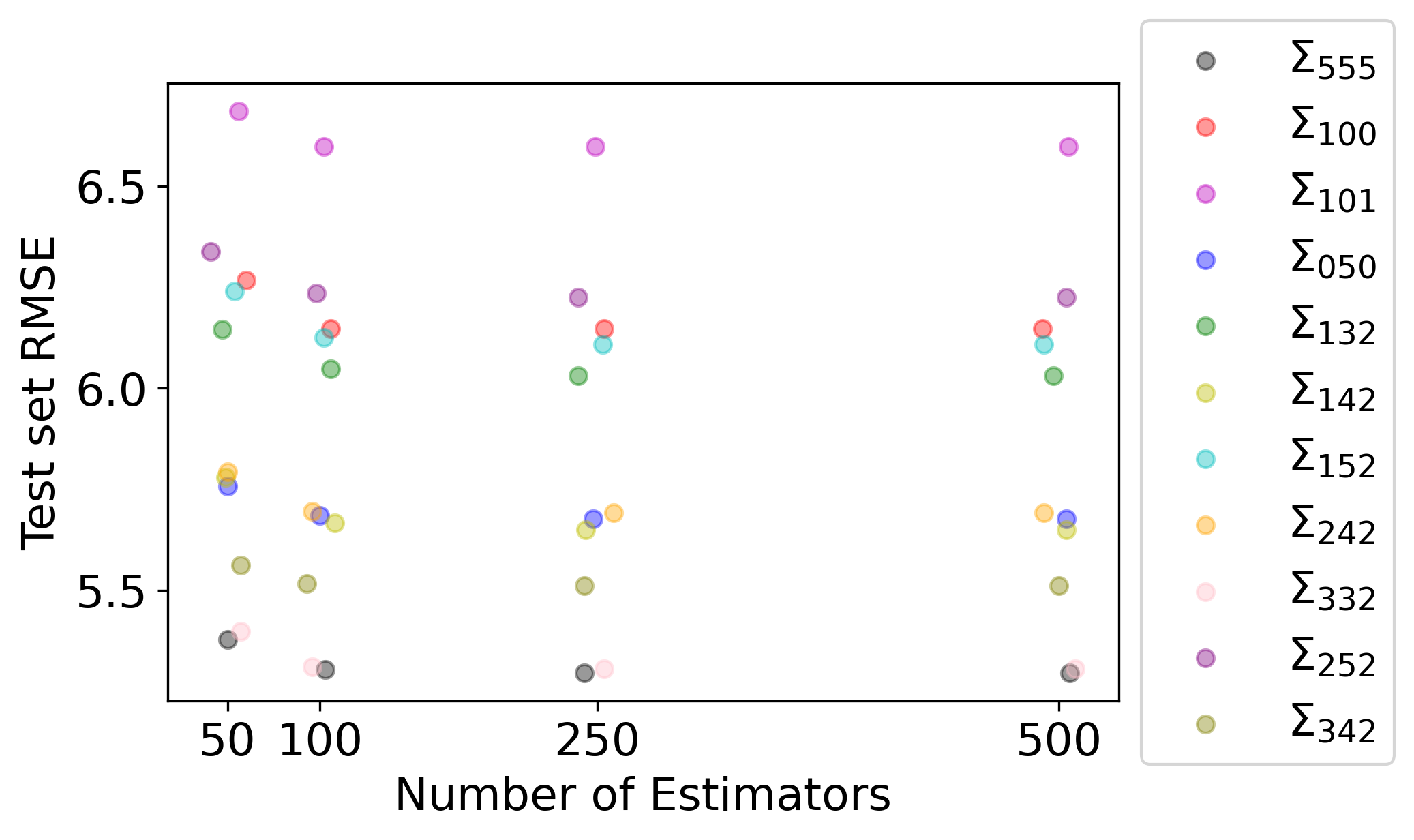}}
\hfill
\caption{
Average (a) MAE and (b) RMSE scores for the test set of formation energy training on a subset of the LPS dataset. A horizontal jitter is applied to the points to help differentiate global structure descriptors.
}
\label{fig:energies}
\end{figure}

\clearpage
\section{Appendix: Comparison between the whole and limited data sets}
\label{appendix:d}
\begin{figure}[ht]
\centering
\hfill
\subfigure[Whole data set]{\includegraphics[width=6.9cm]{figs/Total_rank_with_all.png}}
\hfill
\subfigure[Limited data set]{\includegraphics[width=6.9cm]{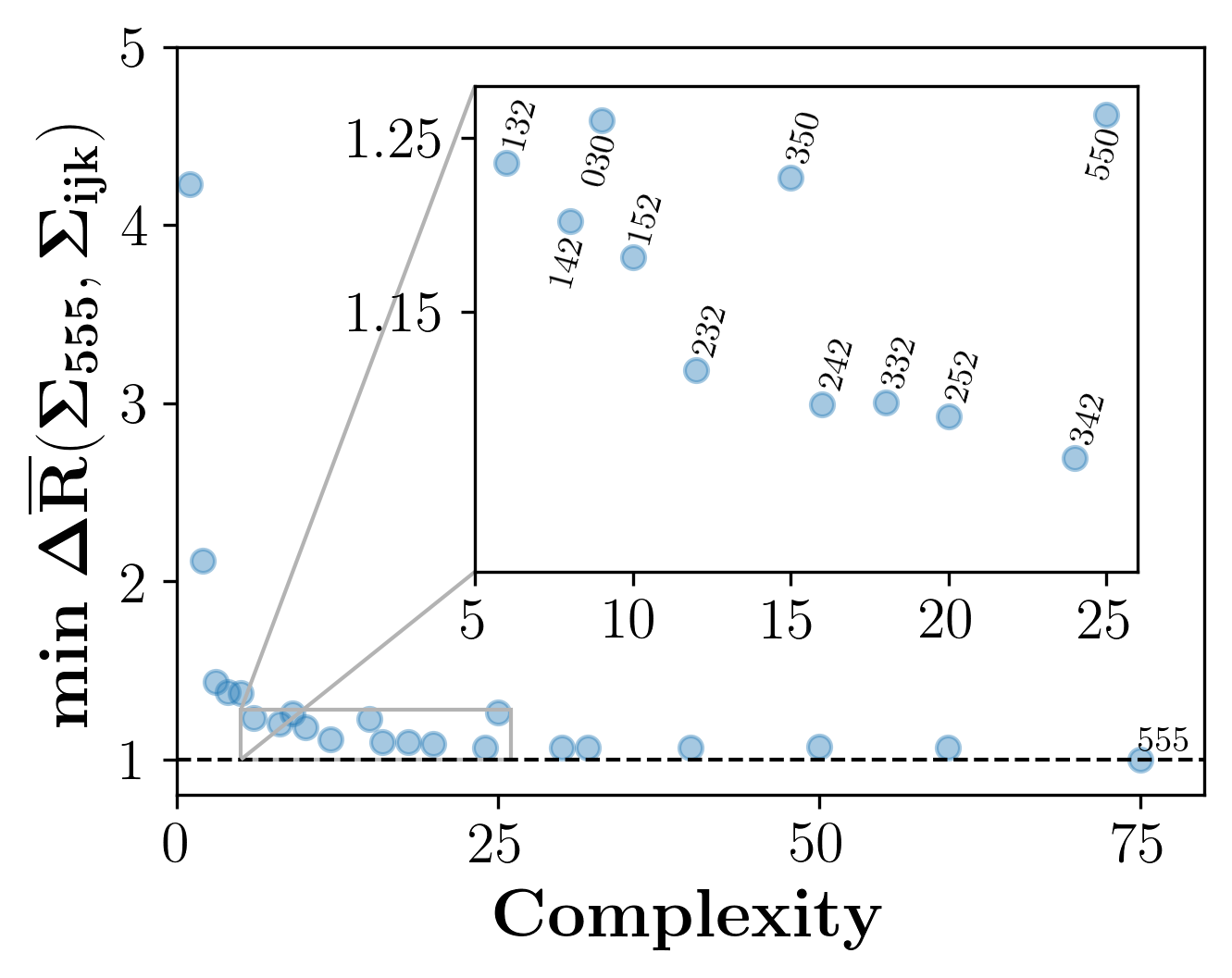}}
\hfill
\subfigure[Whole data set]{\includegraphics[width=6.9cm]{figs/Total_rank_with_all_no_elems.png}}
\hfill
\subfigure[Limited data set]{\includegraphics[width=6.9cm]{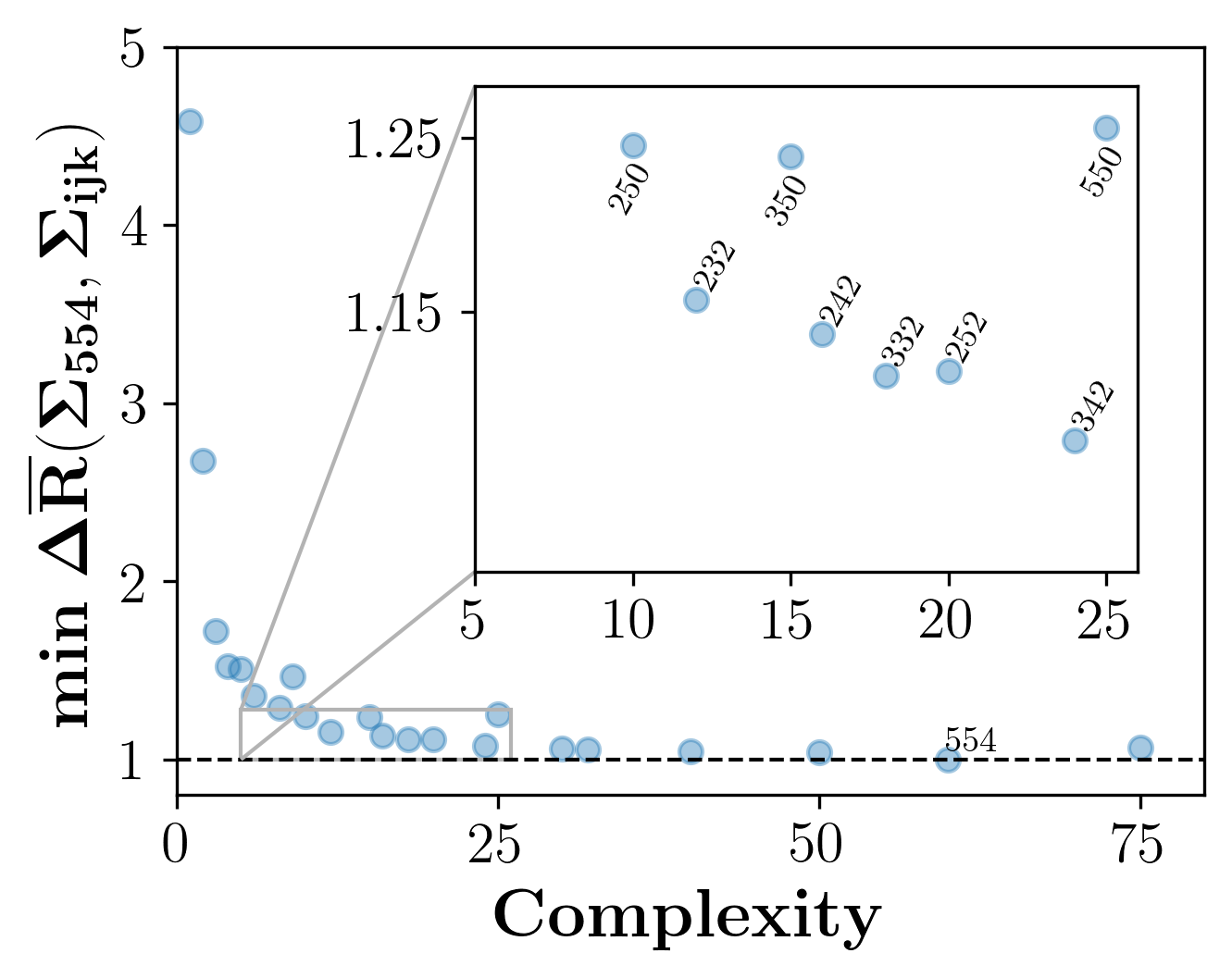}}
\hfill
\caption{
Minimum symmetric rank information imbalance between (a,b) $\Sigma_{555}$, (c,d) $\Sigma_{554}$ and all other descriptors as a function of the descriptor complexity for (a,c) the whole and (b,d) limited data sets.
}
\label{fig:new and old database}
\end{figure}

\begin{figure}[ht]
\centering
\hfill
\subfigure[Whole data set]{\includegraphics[width=6.9cm]{figs/1_.png}}
\hfill
\subfigure[Limited data set]{\includegraphics[width=6.9cm]{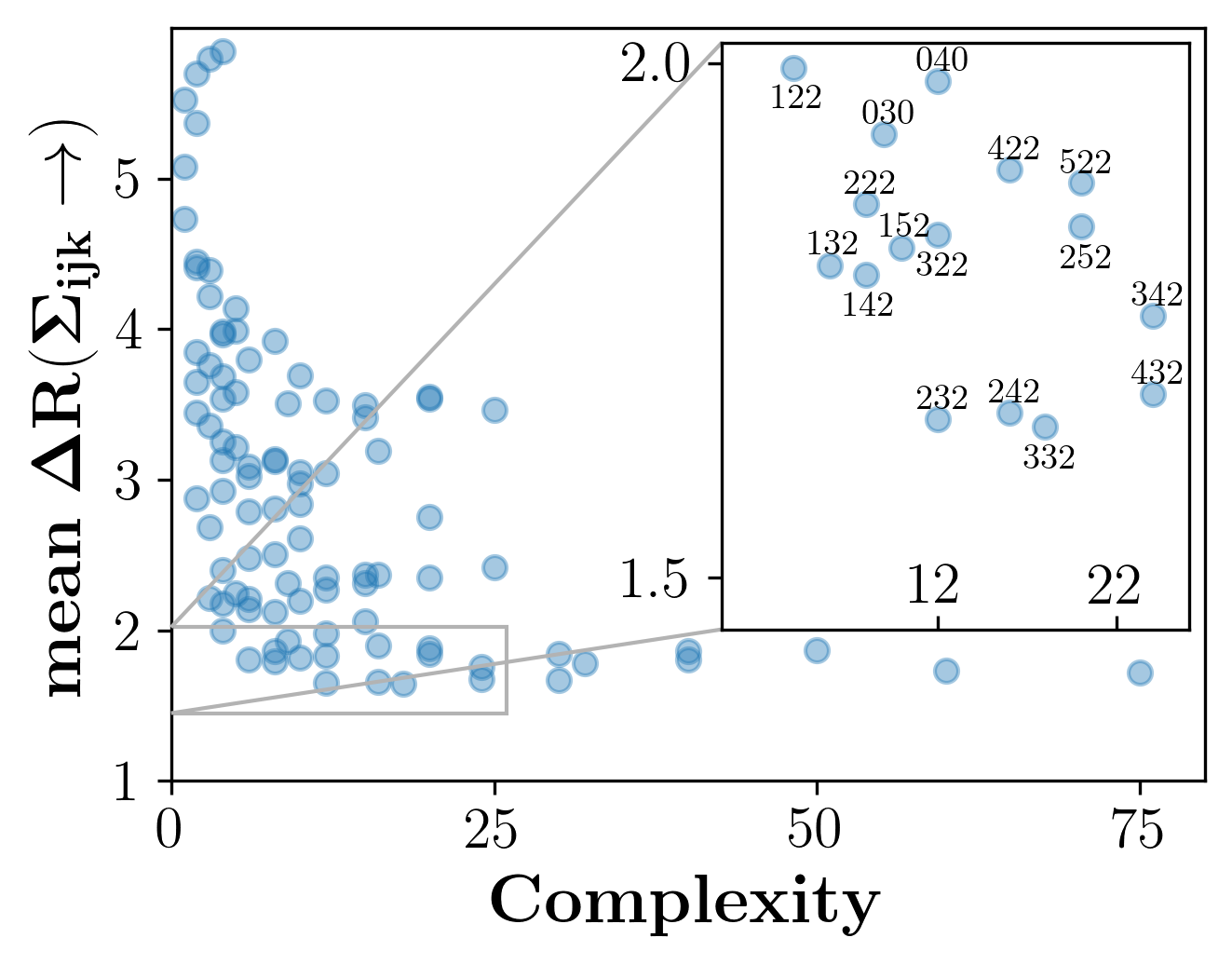}}
\hfill
\subfigure[Whole data set]{\includegraphics[width=6.9cm]{figs/_2.png}}
\hfill
\subfigure[Limited data set]{\includegraphics[width=6.9cm]{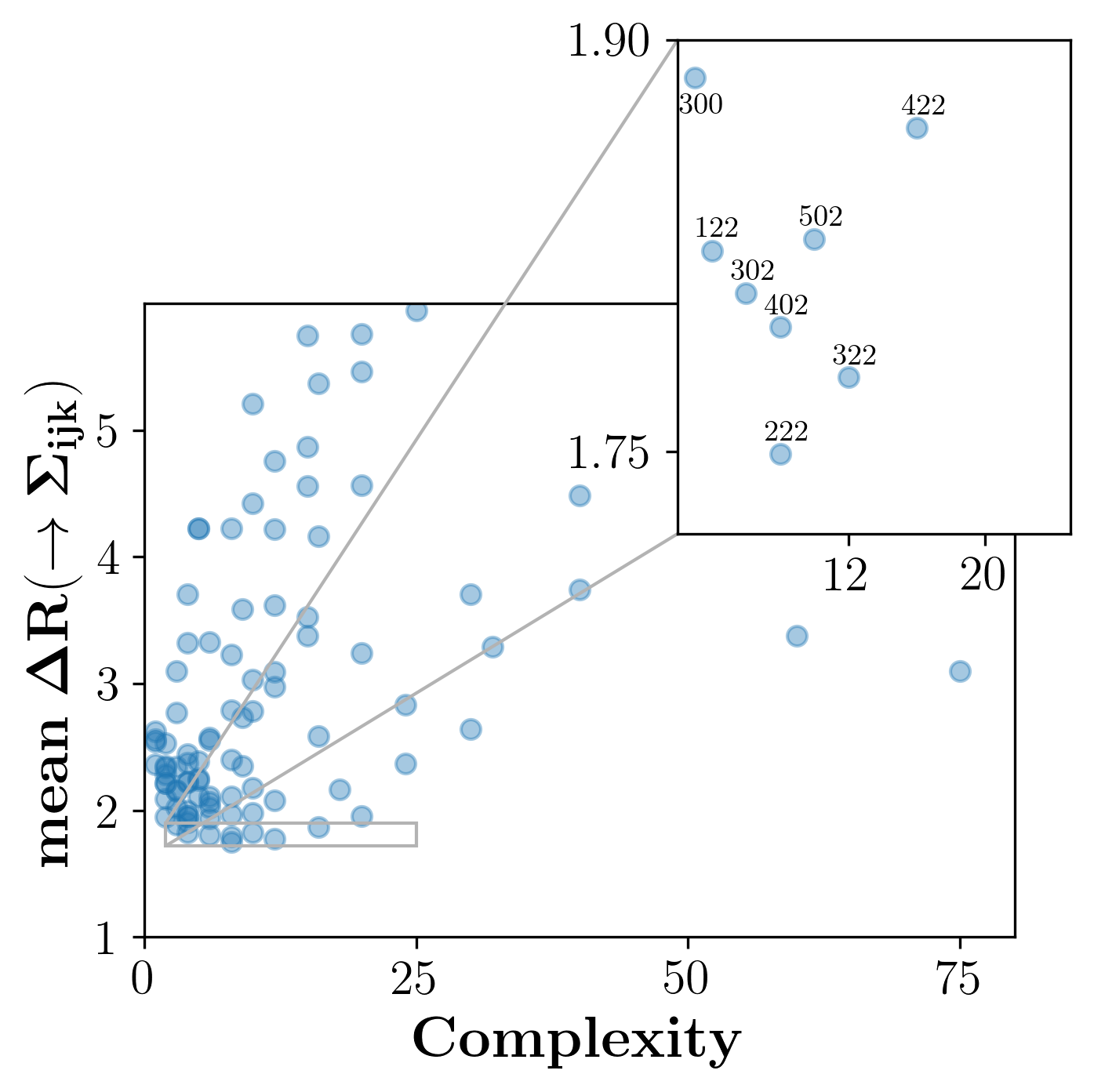}}
\hfill
\caption{
Mean rank information imbalance of inferring all other descriptors from the given descriptor with the given complexity for (a) the whole and (b) limited data sets.
Mean information imbalance of inferring the given descriptor with the given complexity from all other descriptors for (c) the whole and (d) limited data sets.
}
\label{fig:new and old main database}
\end{figure}

\begin{figure}[ht]
\centering
\hfill
\subfigure[Whole data set]{\includegraphics[width=6.9cm]{figs/all_imbalance.png}}
\hfill
\subfigure[Limited data set]{\includegraphics[width=6.9cm]{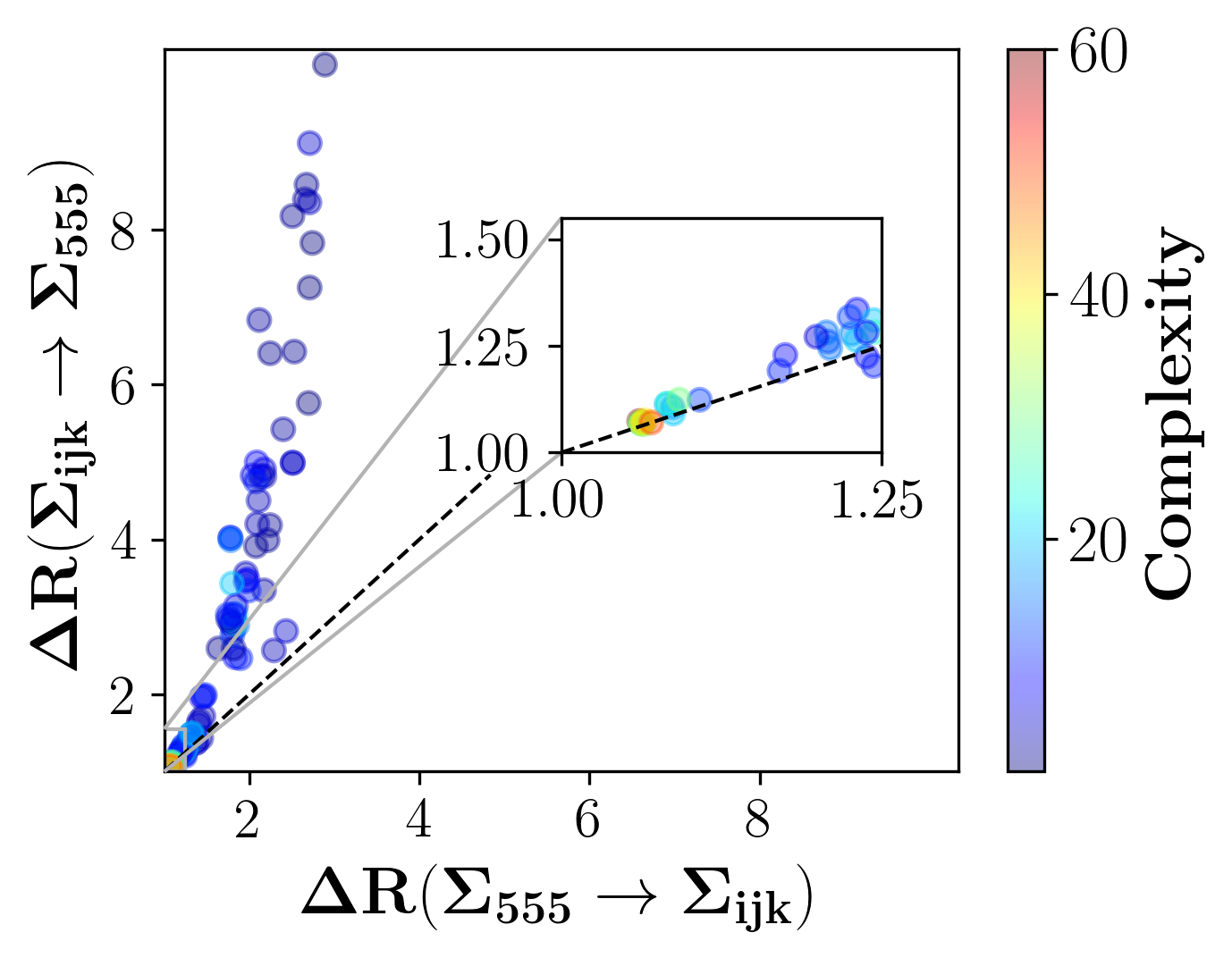}}
\hfill
\subfigure[Whole data set]{\includegraphics[width=6.9cm]{figs/all_no_elems_imbalance.png}}
\hfill
\subfigure[Limited data set]{\includegraphics[width=6.9cm]{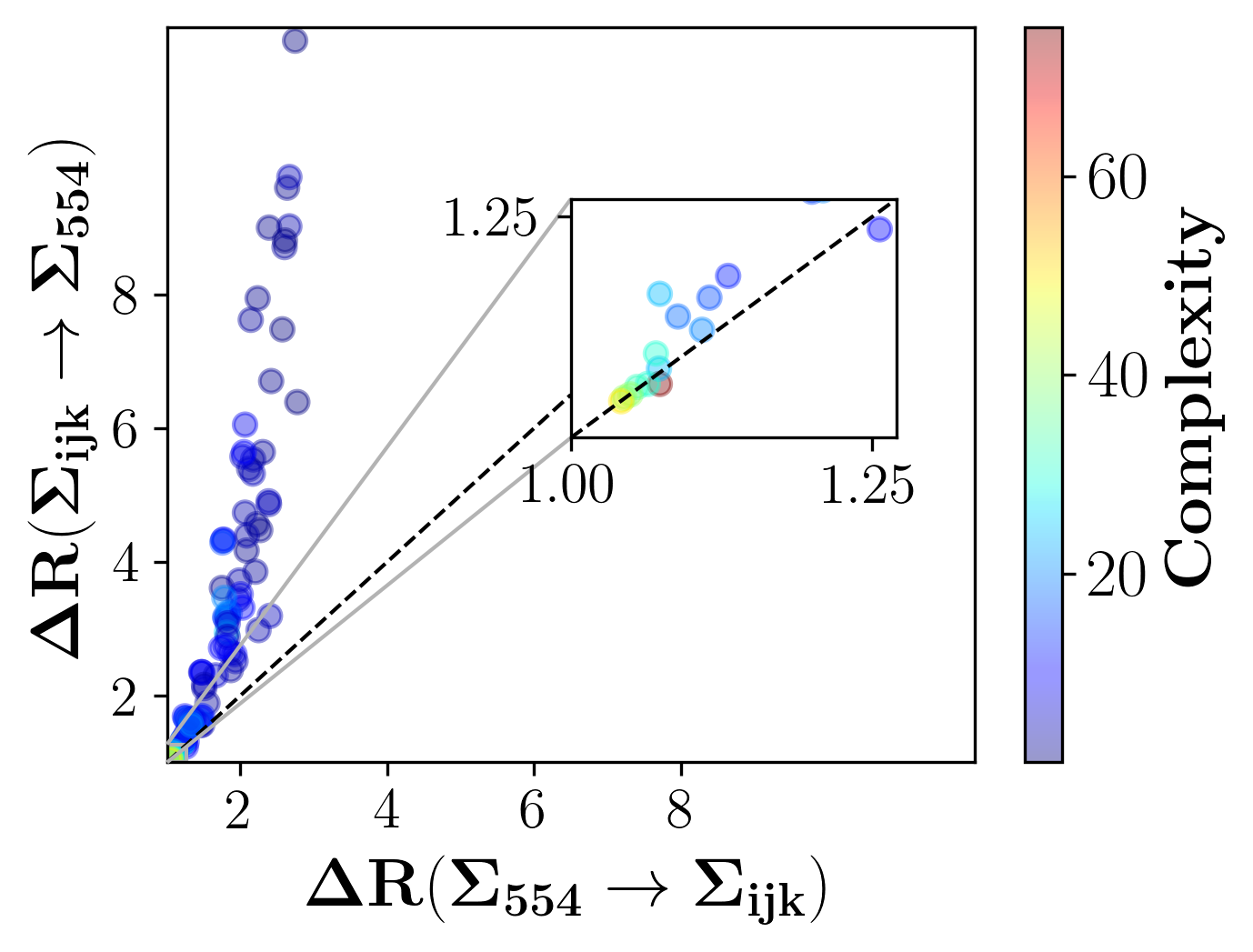}}
\hfill
\caption{
Correlation plot between rank information imbalances of inferring all other descriptors from (a,b) $\Sigma_{555}$, (c,d) $\Sigma_{554}$ (on $x$-axis) and inferring (a,b) $\Sigma_{555}$, (c,d) $\Sigma_{554}$ from all other descriptors (on $y$-axis) for (a,c) the whole and (b,d) limited data sets.
}
\label{fig:new and old main database-correlation}
\end{figure}

In the main body of the paper, we have performed the energy fittings on a smaller data set of structures (compared to the dataset used for the information-theoretic analysis). Here, we compare the results from the information-theoretic study on the whole and limited data sets. From Figures~\ref{fig:new and old database},~\ref{fig:new and old main database}, and~\ref{fig:new and old main database-correlation}, the choice of the data set (in our case slightly) affects the relative ranking of optimally compressed descriptors. For this reason, a similar analysis should first be performed for a new data set before choosing an optimally compressed descriptor.

\clearpage
\section{Appendix: A linear model for formation energy predictions}
\label{appendix:e}
\begin{figure}[ht]
\centering
{\includegraphics[width=9cm]{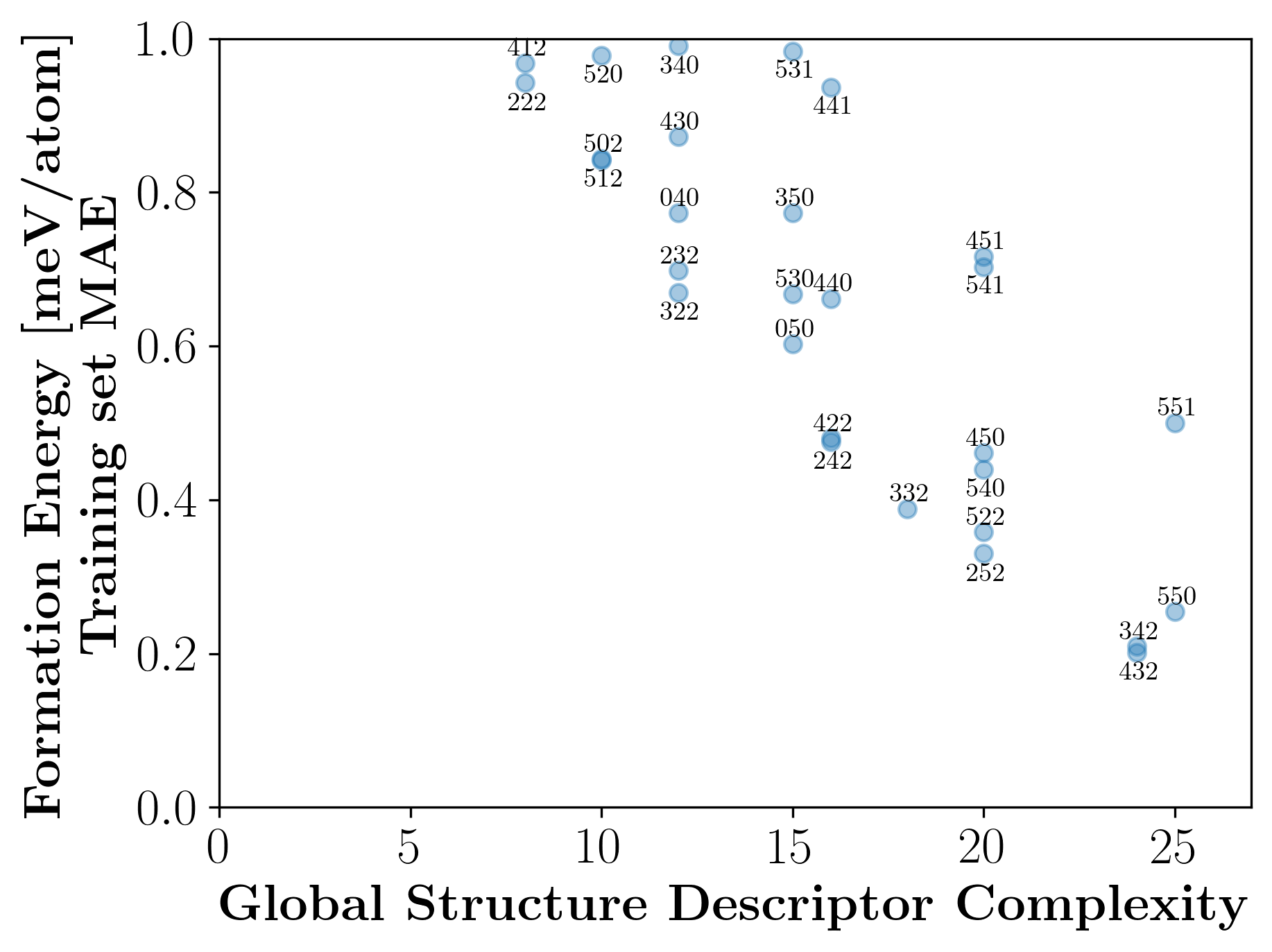}}
\caption{
MAE score of a simple linear regression model trained on the entirety of the limited data set as a function of descriptor complexity. Note that only a small number of relevant descriptors are depicted.
}
\label{fig:linear model}
\end{figure}

We also fitted a linear model to the entirety of the limited data set (the training set contains the entirety of the limited data set).
The relative ordering of descriptors from Figures~\ref{fig:energies}, ~\ref{fig:new and old database}, and~\ref{fig:linear model} is slightly different. 
Note that with the information-theoretic approach, we aim to construct a universal GSD instead of one only useful for the formation energy prediction task. So similar model analyses for other properties are also needed before choosing the universal optimally compressed global descriptor.

\end{document}

%% file: math_commands.tex
\usepackage{amsmath,amsfonts,bm}

\def\eqref#1{equation~\ref{#1}}

\def\1{\bm{1}}

\DeclareMathAlphabet{\mathsfit}{\encodingdefault}{\sfdefault}{m}{sl}
\SetMathAlphabet{\mathsfit}{bold}{\encodingdefault}{\sfdefault}{bx}{n}

\def\sS{{\mathbb{S}}}

\def\sX{{\mathbb{X}}}

